\documentclass[twocolumn,fleqn]{article}

\begin{document}
\twocolumn[
{\bf
SYMMETRIES OF FUNDAMENTAL INTERACTIONS IN QUANTUM PHASE SPACE     
}

\bigskip

{\large\bf V.V. Khruschov}

\medskip

{\it Centre for Gravitation and Fundamental
Metrology, VNIIMS, 46 Ozyornaya St.,
Moscow 119361, Russia
}

\medskip


\vspace*{-0.5cm}

\begin{center}
\begin{abstract}
\parbox{15cm}{Quantum operators of coordinates and momentum components 
 of a particle in  Minkowski space-time belong to a
noncommutative algebra and give rise to a quantum phase space.
Under some constraints, in particular, the Lorentz invariance condition, 
the algebra of observables, including  the Lorentz group generators, 
 depends on  additional fundamental physical constants
with the dimensions of mass, length and action. Generalized 
symmetries in a quantum phase space and some consequences
for fundamental interactions of particles are considered.



\noindent{\bf DOI:} 10.1134/S0202289309040069
}

\end{abstract}
\end{center}
\bigskip
\smallskip
]

\noindent\textbf{\large 1. Introduction}

\smallskip

 It is well known that the Poincar\'e symmetry, which is the space-time 
symmetry of the orthodox relativistic quantum field theory (QFT),
originates from the isotropy and  homogeneity of Minkowski space-time and is 
based on observations of macro- and micro-phenomena concerning 
conventional physical bodies and particles.
 For development of a general quantum theory of fundamental interactions 
it would be desirable to examine in detail possible generalized symmetries
 of fields, their interactions and the underlying space. Studies along 
these lines have been carried out  in the context  of both canonical QFT 
\cite{bog,blox,fu} and its modifications. Among these 
modifications there are theories with new fundamental physical constants 
other than the well known ones $c$ and $\hbar$. Beginning with Snyder's 
work \cite{sny}, a theory with a fundamental length has been elaborated 
\cite{gol,kad}. However, in a theory of this sort, the reciprocity between
coordinates and momenta,  proposed by Born  \cite{born},
was broken. This reciprocity  was restored in by Yang \cite{yang}, 
who added a fundamental mass  to the modified theory.

 In the present paper, we investigate the most general noncommutative algebra
for quantum operators of coordinates and momentum components 
 of a particle in  Minkowski space-time under some natural conditions,
 in particular, the Lorentz invariance condition  \cite{lez1, khru1}.
The general algebra depends on the constants with the dimensions of length
\cite{sny}, mass \cite{yang},  and action \cite{lez1}.
 We consider new generalized symmetries in a phase space and some consequences
for fundamental interactions of particles.

 The  paper  is organized as follows:  Section 2 is
devoted to a brief description of results obtained previously for 
the symmetries of a quantum phase space. Some applications
of these symmetries to fundamental interactions of particles 
are presented in  Section 3. In  Section 4, new generalized symmetries 
in a phase space and some possible consequences for the description of 
particle properties and their interactions are considered. 
Conclusions and a discussion are found in  Section 5.

\smallskip

\noindent{\large\bf 2. Generalized algebra of quantum theory observables}

\smallskip

     We consider a generalized algebra $g$ of quantum theory observables,
when coordinates and momenta are on equal terms and form a generalized 
quantum phase space. The algebra $g$  generated by the  observables can 
depend on extra fundamental constants other than the well 
known $c$ and $\hbar$ \cite{lez1, khru1}. To restrict  the considerable list
of  symmetries corresponding to the algebra $g$,
the following natural conditions are imposed:

  a) The generalized algebra $g$ of observables should be a  Lie algebra;

    b) The  dimension of $g$ should coincide with the dimension of
the algebra of canonical quantum theory observables in Minkowski space-time;

    c) Physical dimensions of the observables, representing the 
generators of $g$, should be the same as the canonical ones;

    d) The algebra $g$ should contain the Lorentz algebra $l$ as its
subalgebra and  commutation relations of the  generators of $l$
with other generators, should be the same as the canonical ones.

Under these conditions, the most general algebra has been derived,
which depends on new constants with the dimensions of length, mass and action,
and the commutation relations can be presented as ($i, j, k, l = 0, 1, 2, 3$)
\[
[F_{ij}, F_{kl}]=if(g_{jk}F_{il}-g_{ik}F_{jl}+g_{il}F_{jk}-g_{jl}F_{ik}),
\]
\[
[F_{ij}, p_{k}]=if(g_{jk}p_{i} - g_{ik}p_j), 
\]
\[ [F_{ij}, x_k]=if(g_{jk}x_i - g_{ik}x_j),
\]
\[
[F_{ij}, I]=0, \quad [p_i, p_j ]=(if/L^2)F_{ij}, 
\]
\begin{equation}
[x_i, x_j]=(if/M^2)F_{ij}, 
\label{al1}
\end{equation}
\[ [p_i, x_j]=if(g_{ij}I +  F_{ij}/H),
\]
\[
[p_i, I]=if(x_i/L^2 - p_i/H),  
\]
\[ [x_i, I]=if(x_i/H - p_i/M^2).
\]

The first relation specifies the algebra $l$, while 
the second, third and fourth relations specify the tensor
character for the well-known physical quantities. The fifth and sixth
relations lead to noncommutativity of $p$ and $x$. 
The seventh, eighth and ninth relations  generalize
 the Heisenberg relation. The system of relations (1) is written 
in units with $c = 1$ ($c$ is the velocity of light), 
it contains four dimensional parameters: $f$(action), $M$(mass), $L$(lenght), 
and $H$(action). But in the limiting case  $M \to\infty$, $L\to\infty$, 
$H\to\infty$, the system (1) should transform to the system of relations
for the canonical quantum theory, so $f = \hbar$. More generally, 
$f= f(M,L,H)$ and in the limiting case $f(M,L,H)\to\hbar$.

The generalized algebra (\ref{al1}) contains as special cases a great
number of Lie algebras of different symmetry groups. The  condition 
for the algebra (\ref{al1}) to be  semisimple  can be written in the
 form: $(M^2L^2 - H^2)/M^2L^2H^2\ne 0$. If this condition is fulfilled,
$g$ is isomorphic to a pseudoorthogonal algebra $o(p,q), p+q=6$, according to 
 $M$, $L$ and $H$ values.  In other cases it is isomorphic to a direct or a
semidirect product of a pseudoorthogonal algebra and an Abelian
or integrable algebra \cite{lez1,khru1}. Let us display below the ranges of
the  M, L and H parameters, which are matched to the o(3,3),
o(4,2), and o(5,1) algebras. If $sign(H^2- M^2L^2)=-1$, 
$sign(M^2)=sign(L^2)$, or $sign(H^2- M^2L^2)=1$, 
$sign(M^2)=-sign(L^2)$, then $g=o(2,4)$.
If $sign(H^2- M^2L^2)=1$, 
$sign(M^2)=sign(L^2)=1$, then $g=o(1,5)$, if $sign(H^2- M^2L^2)=1$, 
$sign(M^2)=sign(L^2)=-1$, then $g=o(3,3)$.

Irreducible  representations for the pseudoorthogonal algebras 
of rank 3 are determined with the help of eigenvalues of  three Casimir
 operators:
$K_1=\varepsilon_{IJKLMN}F^{IJ}F^{KL}F^{MN}$,
 $K_2=F_{IJ}F^{IJ}$, $K_3=(\varepsilon_{IJKLMN}F^{KL}F^{MN})^2$, 
$I,J,K,L,M,N=0,1,2,3,4,5$.
For instance, the second-order invariant operator $K_2$  in terms
of $I, p, x$, and $F$ can be written in the form:
\[
C_2 = \sum_{i<j}F_{ij}F^{ij}(1/M^2L^2 - 1/H^2) + I^2 +
\]
\begin{equation}
 (x_ip^i + p_ix^i)/H - x_ix^i/L^2 - p_ip^i/M^2.
\end{equation}

The mathematical properties of the generalized algebra (\ref{al1}), 
 have been studied in   \cite{lez1,khru1,mendes,chry}.
Apart  from its mathematical properties, the  algebra $g$ is an object of 
interest in modern physical applications as well. For instance, in 
 \cite{lez2}, it is suggested  to apply the algebra 
(\ref{al1}) in  classical physics on astronomical scales.

\smallskip

\noindent{\large\bf 3.  Some applications 
of the algebra $g$  to fundamental interactions of particles
}

\smallskip

We consider possible applications of the algebra $g$ to quantum
phenomena on microscales \cite{khru1,khru2}. 
In this case it is convenient to use the quantum constants $\kappa= \hbar/H$,
$\lambda = \hbar/M$, $\mu = \hbar/L$ and to write the commutation relations
 (\ref{al1}) dependending on the quantum constants as 
(we use the natural units with $c = \hbar = 1$):
\[
 [p_i, p_j ]=i\mu^2F_{ij}, \quad [x_i, x_j]=i\lambda^2F_{ij}, 
\]
\[
 [p_i, x_j]=i(g_{ij}I + \kappa F_{ij}),
\]
\begin{equation}
[p_i, I]=i(\mu^2x_i - \kappa p_i),  
\label{al3}
\end{equation}
\[ [x_i, I]=i(\kappa x_i - \lambda^2p_i).
\]

In the general case, one may classify Generalized Quantum Fields
(GQF) as the fields which form a space for an irreducible
representation of the algebra $g$ (\ref{al3}). For a pseudoorthogonal 
algebra,  GQF should obey the following equation, among others:
\[
[\sum_{i<j}F_{ij}F^{ij}(\lambda^2\mu^2 - \kappa^2) + I^2 + 
\kappa(x_ip^i + p_ix^i) - 
\]
\begin{equation} 
\mu^2x_ix^i - \lambda^2p_ip^i]\Phi = 0.      
\label{ur1}
\end{equation}

Let us apply the algebra  (\ref{al3}) for a description of color particles,
 such as quarks or gluons.  Then additional constraints arise for the
form of this algebra. Due  $CP-$invariance of strong
interactions we obtain $\kappa = 0$. Here we have used a result
obtained in \cite{lez1} that the commutation relations (\ref{al1})
became $CP-$noninvariant at finite $H$ values.  Moreover, in the presence of 
a nonzero $\lambda$ value  some inconsistencies take place in the quark 
descriptions inside hadrons. Thus we put $\kappa = \lambda = 0$ \cite{khru2}.
 Denoting $\mu$  as $\mu_s$, we obtain the following 
nonzero commutation relations besides of the usual commutation relations with
Lorentz group generators, which we do not write below:
\[  \quad    [p_i, p_j] = i\mu_s^2F_{ij},\quad [p_i, x_j] = ig_{ij}I,
\]
\begin{equation} 
\quad [p_i, I] = i\mu_s^2x_i.
\label{al4}
\end{equation} 

 Nonzero uncertainties  immediately follow from these relations for the 
results of simultaneous measurement of quark momentum components. 
For instance, let $\psi_{1/2}$ be a quark state with a definite value of 
its spin component along the third axis. Consequently, $[p_1, p_2] = 
i\mu_s^2/2$, thus $\Delta p_1 \Delta p_2 \ge \mu_s^2/4$ 
 and if $\Delta p_1 \sim \Delta p_2$, one gets $\Delta p_1 >\mu_s/2$,
$\Delta p_2 >\mu_s/2$. Rough estimations in the framework of the quark 
model indicate that the $\mu_s$ value is about 0.5 GeV. 
To extract a more precise number, one can use a quark equation of
Dirac-Gursey-Lee type \cite{dir,gur,khru2}:
\[ [\,\gamma_i(\,p_0^i +dp_0^kL_k^i + i\mu_s\gamma^i/2) + 
\]
\begin{equation}
2i\mu_sS_{ij}(L^{ij} + S^{ij})\,]\psi = m\psi,
\label{ura}
\end{equation}
\noindent   where $p_0 + dp_0L = p_F$  is the space-time total momentum 
\cite{fro}
$d =\mu_s/m_0$ , $p_0$ and $L$  have the forms of the usual generators of 
translations and  Lorentz transformations in Minkowski space-time, 
and $p_0^2 = m_0^2$. To estimate a $\mu_s$ value with the help of 
a constituent quark mass $m$ and a current quark mass $m_0$  values, 
we use a ground state $\psi_0$ in a meson so the contribution from 
$L\psi_0$ can be neglected. In this way we obtain from Eq.(\ref{ura}) 
the approximate relation: $ m\cong m_0 + 2i\mu_s$.

The constant $\mu_s$ should be pure imaginary and negative to account for 
the well-known inequality $m > m_0$. From the 
correspondence between the ranges of parameters and the
pseudoorthogonal groups written above, one can see that the
algebra under consideration is isomorphic to the algebra of the
$AdS$ group $O(2,3)$ \cite{khru2}.

Let us use  the  values of $m$ obtained in the
independent quark model (IQM) on the basis of the hadron spectroscopy data
\cite{khru3}.
If we pick out from the high-energy physics data $m_0\cong 2 MeV$
for the current $u-$quark mass and from the  IQM  $m\cong 316 MeV$, then  we
obtain $|\mu_s|\cong 157 MeV$. Thus one can evaluate the mass values of 
$d-, s-, c-$, and $b-$current quarks on the scale  $ \sim1\, GeV$, which agree 
with  their values obtained in the  QCD framework \cite{ams}.

\smallskip 

\noindent{\large\bf 4.  New 
generalized symmetries in   phase space and their applications
 for description of fundamental particles
}

\smallskip

The pseudoorthogonal groups generated with the help of the quantum
operators of coordinates,  momentum and angular
momentum components  and presented in Section 2,  operate in a 
six-dimensional space.
It should be noted that,  in spite  of numerous attempts, a physical meaning 
of  six-dimensional  coordinates remains unclear. Analogous problems were
considered in Refs.\cite{tol,lez3}. So we use the practice of  embedding
 our O(p,q), p+q=6, groups in more general groups that operate in a
 physical phase space \cite{shch,zach}. 
In this case,  two  pseudoorthogonal groups $O(2,6)$ and $O(4,4)$ arise, and
we argue futher in favour of their  physical significance.

  Actually, as has been described in Section 3 the strong interactions
prefer the $O(2,3)$ symmetry which enters as a part into the 
$O(2,4)$ symmetry, another important physical symmetry (the conformal 
symmetry).  The $O(2,4)$ symmetry is a part of both
 $O(2,6)$ and $O(4,4)$ symmetries. Moreover, if we intend to keep in mind 
the origin of a more general symmetry group, we should take into account the
 three "quantum groups" ($O(2,4)$, $O(1,5)$ and $O(3,3)$) obtained in 
Section 2. The main dificulty consists in a contingency that the $O(2,4)$ 
group belongs  to the $O(2,6)$ and $O(4,4)$ groups while the $O(1,5)$ group 
belongs only to the $O(2,6)$ group, and the $O(3,3)$ group belongs only to 
the $O(4,4)$ group. 

To make a choise between these groups we take into
account result obtained in \cite{khru4} that indicates an association 
of color particles'  confinement  and the possible  $SU(1,3)$ symmetry
of nonperturbative QCD interactions. But the $SU(1,3)$ group lies only in
the $O(2,6)$ group and not in the $O(4,4)$ group. Thus we suggest that, for 
fundamental particles, the generalized mass squared (in the natural units: 
$c=\hbar=1$) has a physical meaning:
\[
df^2 = (dE)^2 -(dp_1)^2 - (dp_2)^2 - (dp_3)^2+
\]
\[
\mu^4(dt)^2-\mu^4(dx_1)^2 -\mu^4(dx_2)^2 -\mu^4(dx_3)^2 =
\]
\begin{equation}
 (dm)^2+\mu^4(ds)^2,
\end{equation}
\noindent where the value of the constant $\mu$ is proportional to 
the $\mu_s$ value. A maximal compact subgroup of the obtained group of 
invariance in the physical phase space is a product of the U(1) and SU(4)
groups, it is important for realizations of representations of the
O(2,6) group or the o(2,6) algebra.

\smallskip

\noindent\textbf{\large 5. Conclusions and discussion}

\smallskip

We have considered in some detail the general "quantum
algebra" of the physical observables which depends on 
additional constants with the dimensions of mass, length and action. 
Generalized symmetries in a quantum phase space and some consequencies
for fundamental interactions of particles have been presented. It 
was argued that the pseudoortogonal O(2,6) symmetry acting in a phase
space has a  physical significance.

When considering the general quantum algebra (\ref{al1}) we mainly restrict 
our treatment of generalized symmetries to the case of strongly interacting 
fundamental particles. This is primarily due to such an unresolved problem 
as  confinement of color particles. Adoption of the O(2,6) and SU(1,3) groups 
for generalized symmetry groups of nonperturbative QCD interactions
has a good potential for solving the confinement problem.

     In  summary,  it  should  be  said that a study  of the
generalized  algebra  (\ref{al1})  and  the properties  of  solutions 
 to Eqs.  (\ref{ur1}) and (\ref{ura}) are important objectives for 
achievement of  mathematical completeness of this approach. A  further 
investigation of the invariance group  in phase space 
 as well as possible physical applications of this invariance is in progress 
 now and will be presented  elsewhere.

\smallskip

\noindent\textbf{\large Acknowledgement}

\smallskip

The author is grateful to V.N. Melnikov, K.A. Bronnikov, V.D. Ivashchuk
and S.V. Bolokhov for useful discussions.

\end{document}